%% file: Main.tex
\documentclass[conference,compsoc]{IEEEtran}

\ifCLASSINFOpdf
\else
\fi
\usepackage{graphicx}
\usepackage{amsthm}
\usepackage{color}
\usepackage{amssymb}
\usepackage{pifont}
\usepackage{multirow, hhline}
\usepackage{subcaption}
\usepackage{scalerel}
\usepackage{tikz}
\usepackage{url}
\usepackage{xurl}
\usepackage{colortbl}
\usepackage[normalem]{ulem}
\usepackage[acronym,shortcuts]{glossaries-extra}
\usepackage[moderate, tracking=normal, leading=normal]{savetrees}

\useunder{\uline}{\ul}{}

\definecolor{blue-back}{rgb}{0.85,0.85,1}
\definecolor{blue-side}{rgb}{0.0,0.3,0.60}
\definecolor{green-back}{rgb}{0.85,1,0.85}
\definecolor{green-side}{rgb}{0.0,0.6,0.30}

\NewDocumentEnvironment{formal-blue}{+b}{%
  \par\smallskip
  \noindent
  \begingroup
  \setlength{\fboxsep}{2.5pt}%
  \colorbox{blue-back}{%
    \hbox{%
      {\color{blue-side}\vrule width 2pt}%
      \hspace{4pt}%
      \begin{minipage}{\dimexpr\linewidth-2\fboxsep-6pt\relax}%
        \setlength{\parindent}{0pt}%
        #1%
      \end{minipage}%
    }%
  }%
  \endgroup
  \par\smallskip
}{}

\NewDocumentEnvironment{formal-green}{+b}{%
  \par\smallskip
  \noindent
  \begingroup
  \setlength{\fboxsep}{2.5pt}%
  \colorbox{green-back}{%
    \hbox{%
      {\color{green-side}\vrule width 2pt}%
      \hspace{4pt}%
      \begin{minipage}{\dimexpr\linewidth-2\fboxsep-6pt\relax}%
        \setlength{\parindent}{0pt}%
        #1%
      \end{minipage}%
    }%
  }%
  \endgroup
  \par\smallskip
}{}

\usepackage[acronym,shortcuts]{glossaries-extra}
\glssetcategoryattribute{acronym}{nohyper}{true}
\setabbreviationstyle[acronym]{long-short}

\newacronym{b-lstm}{BiLSTM}{Bidirectional LSTM}
\newacronym{cdf}{CDF}{Cumulative Distribution Function}
\newacronym{cnn}{CNN}{Convolutional Neural Network}
\newacronym{conditional}{\texttt{Cond-Bin}}{Conditional Previous-Bin}
\newacronym{dl}{DL}{Deep Learning}
\newacronym{empirical}{\texttt{Emp-Pair}}{Empirical Pairs}
\newacronym{ft}{FT}{\textit{Flight Time}}
\newacronym{gan}{GAN}{Generative Adversarial Network}
\newacronym{ht}{HT}{\textit{Hold Time}}
\newacronym{hid}{HID}{Human Interface Device}
\newacronym{lbmc}{LBMC}{Linguistic Buffer and Motor Control}
\newacronym{lstm}{LSTM}{Long Short Time Memory}
\newacronym{ml}{ML}{Machine Learning}
\newacronym{non-stationary}{\texttt{NS-Hist}}{Non-Stationary Histogram}
\newacronym{prng}{PRNG}{Pseudorandom Number Generator}
\newacronym{rq}{RQ}{Research Question}
\newacronym{finding}{RF}{Research Finding}
\newacronym{rf}{RF}{Random Forest}
\newacronym{svm}{SVM}{Support Vector Machine}
\newacronym{vk}{VK}{Virtual Key}
\newacronym{wgan-gp}{WGAN-GP}{Wasserstein GAN with Gradient Penalty}

\begin{document}

\title{QUACK! Making the (Rubber) Ducky Talk:\\A Systematic Study of Keystroke Dynamics for HID Injection Detection}


\author{
\IEEEauthorblockN{
Alessandro Lotto,
Francesco Marchiori,
and Mauro Conti
}
\IEEEauthorblockA{
University of Padua\\
Padua, Italy\\
Email: \{alessandro.lotto@math.unipd.it,
francesco.marchiori@unipd.it,
mauro.conti@unipd.it\}
}
}


\maketitle

\begin{abstract}
Modern computing systems implicitly trust human input devices, allowing USB Human Interface Device (HID) emulators, such as the USB Rubber Ducky, to inject arbitrary keystrokes while bypassing conventional defenses.
Speed- and regularity-based heuristics are easily evaded through slower or randomized timing.
Keystroke dynamics offers a behavioral alternative, but prior work primarily addresses user authentication rather than user-agnostic detection of automated input.
Moreover, detectors that process or retain key identities may expose sensitive content.

In this paper, we investigate whether timing alone can distinguish human from machine-generated keystrokes without per-user enrollment or semantic input features.
We systematically evaluate lightweight classifiers using only hold and flight times in a controlled offline setting.
Across progressively structured synthetic attacker families, we examine cross-generator transfer, mixed-generator training, attacker sophistication, and the effect of observation-window length on discrimination performance and decision delay.
Our results show that lightweight models can effectively discriminate human from machine-generated input across the evaluated generators without content access or user profiling.
Our analysis also suggests that greater synthesis sophistication does not monotonically improve evasion.
Moreover, detection performance primarily depends on exposure to structurally distinct generation strategies rather than model complexity.
This finding can reduce the number of attacker models required for training and supports the feasibility of content-independent, user-agnostic HID-injection detection, while physical HID validation remains necessary.
\end{abstract}


%
\IEEEpeerreviewmaketitle

\input{Sections/1-Introduction}
\input{Sections/2-RelatedWorks}
\input{Sections/3-SystemThreatModel}
\input{Sections/4-Methodology}
\input{Sections/5-ExperimentalEvaluation}
\input{Sections/6-Discussion}
\input{Sections/6.1-Limitations}
\input{Sections/7-Conclusion}

\newpage


\bibliographystyle{IEEEtran}
\bibliography{Bibliography}

\newpage

\appendices
\input{Appendix/Specs}

\end{document}

%% file: Sections/1-Introduction.tex
\section{Introduction} \label{sec:Introduction}
Modern computing systems implicitly trust human input devices, such as keyboards, and grant them direct access to the operating-system input pipeline~\cite{Farhi-et-al,USBattacks,FeasibilityUSBattacks}.
USB \ac{hid} emulators exploit this trust by impersonating keyboards and injecting pre-programmed keystrokes once connected to a target machine.
Tools such as the USB Rubber Ducky~\cite{Hak5_RubberDucky} have consequently evolved from penetration-testing instruments into mechanisms used in cybercrime and espionage campaigns~\cite{USBattacks,FeasibilityUSBattacks,RubberDucky-attack1,RubberDucky-attack2}.
Because injected events traverse legitimate \ac{hid} interfaces, they can execute malicious commands without exploiting software vulnerabilities or triggering conventional malware defenses~\cite{USBattacks,Arora-et-al}.
Existing countermeasures generally detect unusually fast or regular typing~\cite{Arora-et-al,Jothi-et-al}, but attackers can evade these heuristics by slowing the injection or randomizing inter-keystroke delays~\cite{RubberDucky-payloads}.
Effective detection must therefore distinguish human from automated input beyond simple speed and regularity patterns.

Keystroke dynamics provides a behavioral basis for this distinction by modeling features such as key \ac{ht}, i.e., how long a key remains pressed, and inter-keystroke \ac{ft}, i.e., the interval between releasing one key and pressing the next.
These features have been widely used for user authentication and continuous verification~\cite{Gaines-et-al,Killourhy-et-al,Monrose-et-al}.
However, authentication and injection detection optimize different objectives.
User-specific authentication determines whether input matches an enrolled identity; rejecting an unseen legitimate user may therefore be correct~\cite{Rahman-et-al,Serwadda-et-al,Stefan-Yao}.
In contrast, a user-agnostic injection detector must accept input from previously unseen humans while rejecting automated sequences without relying on identity claims or per-user enrollment.
User-specific authentication can remain a complementary control where enrollment is practical, but it does not replace population-level injection detection in shared or general-purpose systems.
Adapting existing algorithms to this task changes the class definition, error semantics, and generalization objective.
In fact, the legitimate class spans diverse human behavior, while synthetic generators may introduce strategy-specific artifacts.
Removing the user template also gives the attacker a broader evasion target, because any timing pattern classified as human may suffice rather than the behavior of one enrolled identity.
We therefore evaluate detection across progressively structured attacker models rather than assuming that performance against one generator transfers to others.

\textbf{Research Questions}.
We study user-agnostic discrimination between human- and machine-generated keystrokes using only timing information.
Our goal is an empirical feasibility and sensitivity analysis, not the optimization of a state-of-the-art classifier.
We deliberately use relatively simple detector architectures and fixed configurations to isolate the effects of attacker structure, training composition, and observation-window length.
Throughout the paper, \emph{robustness} denotes transfer across the generator families evaluated under our controlled dataset and offline analysis pipeline; it does not imply robustness against arbitrary physical \ac{hid} implementations or real-world deployment conditions.

A straightforward defense would consist of continuously monitoring user input through a keylogging mechanism integrated with an intrusion detection system to identify anomalous typing behavior.
This, however, raises significant privacy concerns.
Besides, continuous detection must also remain computationally lightweight and avoid the enrollment and maintenance of user-specific profiles.
This introduces our first research question:
\begin{formal-blue}
\textbf{RQ1} -- \textit{Can \ac{hid}-based keystroke injection attacks be reliably detected through continuous keystroke dynamics analysis using lightweight classifiers without per-user enrollment or semantic input features?}
\end{formal-blue}
Attackers can move beyond simple timing randomization and generate increasingly structured temporal behavior.
A detector trained against one strategy may therefore learn generator-specific artifacts rather than general differences between human and automated input.
This motivates evaluating whether detection transfers to previously unseen generation processes:
\begin{formal-blue}
\textbf{RQ2} -- \textit{To what extent do detectors generalize across heterogeneous attacker models employing different statistical keystroke generation strategies?}
\end{formal-blue}
Cross-generator transfer also depends on which attacker models the defender includes during training.
Because exhaustive coverage of possible generators is impractical, we investigate which training composition provides broad detection without requiring every individual strategy:
\begin{formal-blue}
\textbf{RQ3} -- \textit{Which training strategy enables the most robust detection performance against diverse keystroke generation processes?}
\end{formal-blue}
Practical detection must further balance timeliness and reliability.
Short windows enable earlier decisions but provide limited temporal evidence, whereas longer windows may improve discrimination only after more of the injected sequence has executed.
Determining the optimal observation window becomes a key design parameter.
\begin{formal-blue}
\textbf{RQ4} -- \textit{How does the keystroke observation-window length affect the trade-off between discrimination performance and decision delay?}
\end{formal-blue}
Finally, more sophisticated generators may better reproduce human timing dependencies, but greater model complexity does not necessarily yield more effective evasion.
We thus assess whether synthesis sophistication is itself a reliable indicator of attack effectiveness:
\begin{formal-blue}
\textbf{RQ5} -- \textit{Does increasing the statistical sophistication of synthetic keystroke generation significantly improve an attacker's ability to evade detection?}
\end{formal-blue}

\textbf{Contributions}.
This study makes the following contributions:
\begin{enumerate}
    \item We present a systematic, threat-model-driven evaluation of user-agnostic \ac{hid} injection detection across naive randomization, context-aware statistical generation, and adaptive neural generation, using only \ac{ht} and \ac{ft} and requiring neither per-user enrollment nor semantic input features.

    \item Cross-generator and mixed-generator evaluation show that lightweight detection generalizes primarily through coverage of structurally distinct attacker families rather than exhaustive coverage of individual generators or increased detector complexity, thereby reducing training requirements.

    \item We quantify how observation-window length affects discrimination performance and delay, and we measure the inference cost of representative lightweight configurations.

    \item We release our experimental pipeline in a GitHub repository~\footnote{\url{https://anonymous.4open.science/r/QUACK-8447}} (anonymized for submission) to support reproducibility and future comparative studies.
\end{enumerate}

\textbf{Organization.}
The remainder of the paper is organized as follows.
Sec.~\ref{sec:RelatedWork} discusses the current literature in keystroke dynamic analysis.
Sec.~\ref{sec:Model} presents the considered system and threat models.
Sec.~\ref{sec:Methodology} details the systematic evaluation methodology.
Sec.~\ref{sec:Evaluation} presents the experimental evaluation and results, while Sec.~\ref{sec:Discussion} provides detailed analysis, results interpretation and takeaways.
Sec.~\ref{sec:Limitations} highlights limitations and future work, and finally, Sec.~\ref{sec:Conclusion} concludes the paper.

%% file: Sections/2-RelatedWorks.tex
\section{Related Work} \label{sec:RelatedWork}
Keystroke dynamics is an established behavioral biometric that uses temporal motor patterns, particularly key \ac{ht} and inter-key \ac{ft}, for user authentication and anomaly detection~\cite{Gaines-et-al,Killourhy-et-al,Monrose-et-al}.
Early evaluations, however, commonly assumed "zero-effort" human impostors rather than adversaries that actively imitate legitimate timing.
Studies of synthetic forgeries showed that even simple imitation can substantially increase false acceptance rates, exposing the limits of this assumption~\cite{Rahman-et-al,Mhenni-et-al}.
Initial attacks such as NoiseBot and GaussianBot sampled Uniform or Gaussian timings, optionally with first-order Markov dependence~\cite{Stefan-Yao}.
Later work analyzed vulnerabilities across larger datasets~\cite{Serwadda-et-al} and found that heavy-tailed distributions can better approximate empirical timing~\cite{Migdal-et-al}, while the \ac{lbmc} model combined log-normal timing with hidden-state modeling to capture cognitive effects~\cite{Monaco-et-al}.
\textit{González et al.}~\cite{Gonzalez-et-al} further introduced higher-order context and empirical \ac{cdf}-based generators, together with a second-stage liveness detector that identifies forgeries whose timing distributions appear excessively smooth.
Their results also demonstrate the greater threat posed by attackers with access to subject-timing information.

Despite this progression, most keystroke-dynamics research remains framed as user authentication.
Accordingly, the system determines whether input matches a specific enrolled user or a forgery of that user's behavior~\cite{Stefan-et-al}.
Existing methods can detect synthetic or replayed input, but often retain user-specific assumptions, consider only a limited set of generation strategies, or require sensing and hardware capabilities unavailable on conventional keyboards~\cite{Farhi-et-al,Hazan-et-al,Stanciu-et-al}.
They therefore do not establish how timing-based detection generalizes in a user-agnostic \ac{hid}-injection setting, where previously unseen legitimate users must be accepted without enrollment.

Security research has examined \ac{hid}-emulation attacks in parallel.
\textit{Arora et al.}~\cite{Arora-et-al} detected abnormally high typing speeds, while USB Rubber Ducky Hunter combined device fingerprinting with input-frequency monitoring~\cite{Jothi-et-al}.
Such defenses address naive high-speed injections but can be evaded by throttling input or adding timing variability.
\textit{Negi et al.}~\cite{Negi-et-al} applied keystroke dynamics to free-text injection detection, although their evaluation relied primarily on classical statistical assumptions and did not systematically examine higher-order attacker models.
More recent work has used \acp{gan} for realistic keystroke synthesis and defensive augmentation.
The \ac{wgan-gp} framework provides stable adversarial sequence generation~\cite{Gulrajani-et-al}, while BeCAPTCHA~\cite{DeAlcala-et-al} and USB-GATE~\cite{Chillara-et-al} use \ac{gan}-generated samples to improve detection robustness.
These studies establish the value of generative augmentation, but do not systematically characterize transfer among independent randomization, context-aware statistical models, and learned generators.

Prior work therefore establishes that timing can distinguish human input from synthetic forgeries~\cite{Gonzalez-et-al,Stefan-et-al}, including in USB-injection scenarios~\cite{Negi-et-al}.
Our contribution is not the initial formulation of machine-generated input as a detection class, but its systematic evaluation under four unresolved conditions.
First, an \textbf{evaluation-framing gap} remains because a general-purpose detector must accept unseen legitimate users while rejecting automated input without identity or enrollment.
Second, a \textbf{deployment-efficiency gap} concerns whether lightweight models can generalize sufficiently for continuous operation.
Third, prior evaluations do not provide \textbf{systematic adversarial escalation} across independent randomization, context-aware statistical generation, and learned generators, making general detection difficult to separate from generator-specific artifacts.
Finally, the \textbf{trade-off between observation length and detection reliability} remains insufficiently characterized.
We address these gaps through a unified evaluation of attacker structure, cross-generator transfer, training composition, detector complexity, and observation-window length.

%% file: Sections/3-SystemThreatModel.tex
\section{System and Threat Model} \label{sec:Model}

\begin{figure*}[!t]
    \centering
    \includegraphics[width=.75\textwidth]{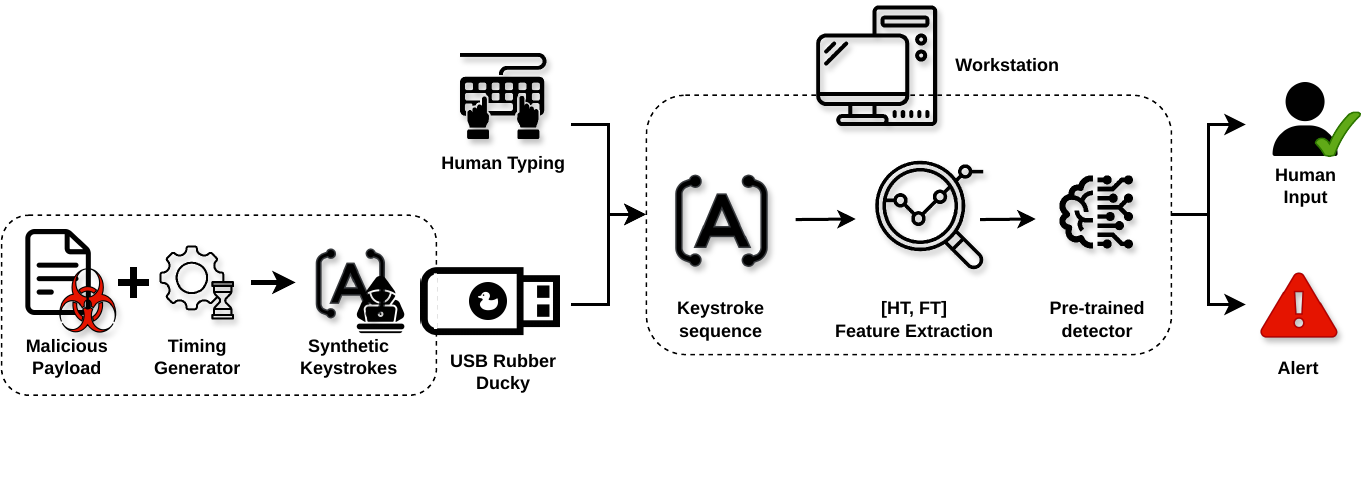}
    \caption{System and threat model. A malicious USB \ac{hid} injects automatically generated keystrokes. The endpoint locally extracts timing features and applies a pre-trained classifier to distinguish human from machine-generated input.}
    \label{fig:SystemModel}
\end{figure*}

\textbf{System Model.}
Fig.~\ref{fig:SystemModel} illustrates the considered system.
The target endpoint accepts keystrokes from a conventional keyboard or any USB device that the operating system enumerates as a keyboard-class \ac{hid}.
A lightweight passive detector monitors the resulting operating-system event stream in real or near-real time, extracts \ac{ht} and \ac{ft} from limited-length sequences.
Thus, it classifies each sequence as human- or machine-generated, raising an alert for suspicious machine-generated input.
We focus exclusively on behavioral timing and exclude device fingerprinting and hardware modifications.
The defender trains the detector offline using measured human sessions and synthetic attack sessions generated under attacker models defined.
The detector is \textit{user-agnostic}, meaning it requires neither per-user enrollment nor long-term profiling and must generalize to legitimate users absent from its training data.
It is also \textit{content-independent} at the classifier boundary.
Although the operating system necessarily receives key events, the classifier receives only the derived \ac{ht} and \ac{ft} sequences, not key identities or reconstructed text.
An intended deployment computes these features locally and immediately discards key identifiers, without retaining or transmitting semantic input content.

\textbf{Threat Model.}
We consider an adversary who causes a programmable malicious \ac{hid} to be connected to an endpoint that accepts it as a keyboard.
We do not require the attacker to have prior control of a physical keyboard.
The attacker may obtain brief access to an exposed USB port, induce a legitimate user to connect an apparently benign device, or plant a malicious keyboard, cable, or peripheral before execution.
These scenarios are consistent with demonstrated BadUSB and USB keypress-injection attacks~\cite{USBattacks,Neuner2018USBlock}.
They also occurred in victim-assisted FIN7 campaigns targeting transportation, insurance, and defense organizations: malicious USB devices were mailed to employees and, once inserted, registered as keyboards and issued PowerShell commands to install malware and support subsequent ransomware operations~\cite{FIN7USB}.
Programmable \acp{hid} remain relevant even when some physical proximity is possible.
Their advantage is not limited to raw typing speed.
They convert seconds of port access, a planted peripheral, or a victim's later insertion into deterministic payload execution.
Compared with manual typing, this reduces attacker dwell time and visibility, avoids transcription errors, and enables repeatable deployment across targets.
The attacker may also deliberately slow or randomize the sequence while retaining these operational advantages.
Moreover, some industrial consoles, unattended terminals, and appliances typically lack an attached keyboard but retain an accessible USB interface that accepts keyboard input for operation or maintenance.
Connecting a programmable \ac{hid} supplies an automated input interface to such an endpoint without requiring an existing keyboard.

The attack nevertheless requires a security-relevant interaction context.
The endpoint must accept the device and map its keystrokes to an actionable interface, such as an active authenticated session, a terminal or interface that accepts shortcuts, text commands, or useful pre-authentication functionality.
A Rubber Ducky does not inherently bypass traditional authentication.
If the endpoint is locked and exposes no security-relevant pre-authentication interaction, the attack considered here cannot proceed.

The attacker controls the injected payload and the timing of every keystroke, but neither knows the detector parameters nor can disable or modify the detector.
Because detection is user-agnostic, evasion does not require imitating a particular enrolled user, but any timing sequence classified as human suffices.
The attacker may therefore slow the input, add noise, sample empirical timing data, or learn temporal dependencies from human sessions.
We evaluate three capability levels.
A \textit{naive attacker} uses independent randomization or lightweight structure-aware sampling and serves as an intentionally weak baseline.
A \textit{context-aware statistical attacker} conditions timing on preceding keystrokes to preserve higher-order dependencies.
An \textit{adaptive attacker} uses \ac{gan} models to approximate human-like timing sequences.
The latter two models represent the principal adversaries and test whether detection remains effective when the attacker explicitly learns temporal structure from human data.

%% file: Sections/4-Methodology.tex
\section{Methodology}\label{sec:Methodology}
\begin{figure*}[!htbp]
    \centering
    \includegraphics[width=.75\textwidth]{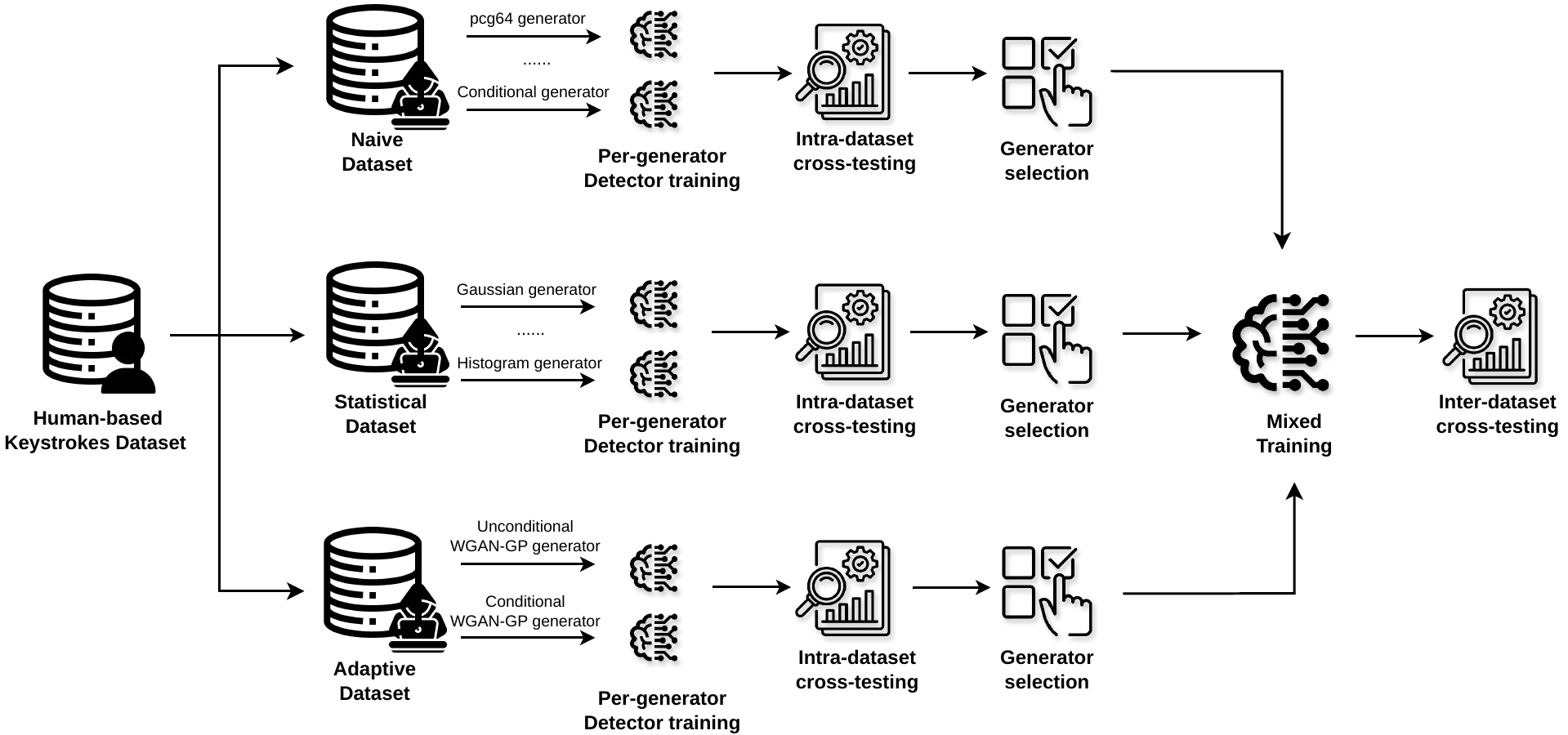}
    \caption{Methodology pipeline for progressive adversarial detector evaluation.}
    \label{fig:Methodology}
\end{figure*}

We evaluate how detector performance changes as attacker timing models become progressively more sophisticated.
Fig.~\ref{fig:Methodology} illustrates the three-stage methodology implemented: \textit{(i)} synthetic dataset generation, \textit{(ii)} single-generator training and cross-testing, and \textit{(iii)} mixed-generator training and evaluation.

\textbf{Dataset Provenance and Generation}. \label{sec:SyntheticGen}
We use the public corpus released by \textit{Gonzalez et al.}~\cite{Gonzalez-et-al}, which consolidates measured human sessions from three independently collected keystroke-dynamics datasets spanning different typing tasks and populations.
Each source session contains \ac{vk} codes, \ac{ht}, and \ac{ft}.
We retain the measured session as legitimate human input and derive paired synthetic sessions by preserving its \ac{vk} sequence while replacing \ac{ht} and \ac{ft} according to the selected attacker model.
We use \ac{vk} codes only during offline construction to align every human session with synthetic counterparts that contain the same text.
\textit{All evaluated classifiers receive only \ac{ht} and \ac{ft}}.
This pairing prevents textual differences from confounding timing-based discrimination.
Accordingly, our content-independence claim applies at the classifier boundary, as the detector does not receive key identities or reconstructed text.

The synthetic sessions model attacker-controlled timing strategies rather than measurements replayed and recaptured through a physical \ac{hid}.
This abstraction reflects that a programmable \ac{hid} schedules key events algorithmically, allowing independent delays, empirical sampling, context-dependent distributions, or learned generation~\cite{Gulrajani-et-al,Mirza-et-al}.
Table~\ref{tab:Datasets} specifies the generators and construction of the three progressively structured dataset families.
The \textbf{Naive Dataset} is an intentionally weak baseline representing attempts to evade speed-based heuristics without modeling realistic temporal structure.
Its \ac{prng}-based generators (\texttt{mt19937}, \texttt{pcg64}, and \texttt{philox}) independently sample timing values from empirical human ranges.
Its two lightweight statistical generators add limited structure: the \ac{empirical} sampler draws joint \ac{ht}-\ac{ft} pairs from the global human distribution, while the \ac{conditional} sampler introduces first-order dependence by conditioning on discretized prior timing states.
The \textbf{Statistical Dataset}, aligned with \textit{Gonzalez et al.}~\cite{Gonzalez-et-al}, uses finite-context generators that condition timing on preceding keystrokes to preserve higher-order dependencies.
It includes the parametric \texttt{Average}, \texttt{Uniform}, and \texttt{Gaussian} generators and the empirical \texttt{Histogram} and \ac{non-stationary} generators.
The \textbf{Adaptive Dataset} represents the more sophisticated attacker through unconditional and context-conditioned \ac{wgan-gp} synthesis~\cite{Gulrajani-et-al,Mirza-et-al}.
The unconditional model generates fixed-length timing windows, whereas the conditional model incorporates recent timing history to approximate local temporal dynamics~\cite{Jinsung-et-al}.
Both learn complex dependencies directly from human data and are optimized to produce realistic timing sequences.

\input{Tables/Datasets}

\textbf{Discriminator Models}.
We compare lightweight non-sequence-aware and sequence-aware detectors.
The former operate on fixed-length representations and comprise \ac{rf} and \ac{svm}; the latter process temporal order and comprise a one-dimensional \ac{cnn}, a unidirectional \ac{lstm}, and a bidirectional \ac{lstm} (BiLSTM).

\textbf{Single-Generator Training Evaluation}.
For each generator, we independently train every detector to distinguish its synthetic sessions from human sessions.
This setting measures separability when the attacker model is known.
We repeat training across multiple input lengths to determine how the number of observed keystrokes affects discrimination.
Input length, therefore, acts as a proxy for the evidence required before classification.
We then cross-test each trained detector against unseen generators within the same generator class, across classes in the same dataset family, and across attacker families.
This analysis distinguishes generator-specific artifacts from transferable human-versus-machine characteristics, identifies challenging and broadly transferable generators, and guides generator selection for mixed training.

\textbf{Mixed-Generator Training Evaluation}.
We next train detectors on mixtures of representative generators selected from the single-generator results.
\textit{Balanced configurations} (BC) assign equal representation to their included generators.
\textit{Unbalanced configurations} (UC) instead emphasize generators that provide broader transfer, testing whether calibrated exposure improves robustness without exhaustive coverage of every strategy.
Training proportions vary across UC configurations, but evaluation always uses the same number of samples from each test dataset.
This separation prevents the training mixture from biasing test prevalence and ensures that reported performance reflects cross-generator generalization rather than majority-class dominance.
Sec.~\ref{sec:Evaluation} specifies the evaluated mixtures and their proportions.

%% file: Tables/Datasets.tex
\begin{table*}[]
\centering
\renewcommand{\arraystretch}{1.35}
\setlength{\tabcolsep}{5pt}
\caption{Synthetic dataset organization.}
\label{tab:Datasets}
\resizebox{.75\textwidth}{!}{%
\begin{tabular}{l|c|c|c|c}
\hline
\rowcolor[HTML]{C0C0C0} 
\multicolumn{1}{c|}{\cellcolor[HTML]{C0C0C0}\textbf{Dataset Family}} &
  \multicolumn{1}{c|}{\cellcolor[HTML]{C0C0C0}\textbf{Generator Class}} &
  \textbf{Generators} &
  \multicolumn{1}{c|}{\cellcolor[HTML]{C0C0C0}\textbf{Approach}} &
  \textbf{Attacker model} \\ \hline
 &
  PRNG &
  mt19937, pcg64, philox &
  \begin{tabular}[c]{@{}c@{}}Independent sampling from\\ global empirical ranges\end{tabular} &
  \begin{tabular}[c]{@{}c@{}}Independent\\randomization\\\end{tabular} \\ \cline{2-5} 
 &
   &
  Conditional PrevBin &
  \begin{tabular}[c]{@{}c@{}}First-order conditional sampling\\ (binned)\end{tabular} &
   \\ \cline{3-4}
\multirow{-5}{*}{\begin{tabular}[c]{@{}l@{}}\textit{Naive}\\ \textit{Dataset}\\(Baseline)\end{tabular}} &
  \multirow{-2}{*}{\begin{tabular}[c]{@{}c@{}}Lightweight\\ statistical\end{tabular}} &
  Empirical Pairs &
  \begin{tabular}[c]{@{}c@{}}Joint pair sampling\\ (HT–FT)\end{tabular} &
  \multirow{-3}{*}{\begin{tabular}[c]{@{}c@{}}Lightweight\\structure-aware.\end{tabular}} \\ \hline
\rowcolor[HTML]{EFEFEF} 
\cellcolor[HTML]{EFEFEF} &
  \begin{tabular}[c]{@{}c@{}}Parametric\\ context-based\end{tabular} &
  Average, Uniform, Gaussian &
  \begin{tabular}[c]{@{}c@{}}Preserve higher order\\  dependencies\end{tabular} &
  \cellcolor[HTML]{EFEFEF} \\ \cline{2-4}
\rowcolor[HTML]{EFEFEF} 
\cellcolor[HTML]{EFEFEF} &
  \cellcolor[HTML]{EFEFEF} &
  Histogram &
  \begin{tabular}[c]{@{}c@{}}Finite-context modeling\\ with empirical distributions\end{tabular} &
  \cellcolor[HTML]{EFEFEF} \\ \cline{3-4}
  
\rowcolor[HTML]{EFEFEF} 
\multirow{-5}{*}{\cellcolor[HTML]{EFEFEF}\begin{tabular}[c]{@{}l@{}}\textit{Statistical}\\ \textit{Dataset}\end{tabular}} &
  \multirow{-2}{*}{\cellcolor[HTML]{EFEFEF}\begin{tabular}[c]{@{}c@{}}Empirical\\ context-based\end{tabular}} &
  Non-Stationary Histogram &
  \begin{tabular}[c]{@{}c@{}}Context modeling +\\ non-stationary offset modeling\end{tabular} &
  \multirow{-5}{*}{\cellcolor[HTML]{EFEFEF}\begin{tabular}[c]{@{}c@{}}Context-aware\\ statistical attacker\end{tabular}} \\ \hline
 &
   &
  Unconditional WGAN-GP &
  \begin{tabular}[c]{@{}c@{}}Direct distribution learning\\ via adversarial training\end{tabular} &
   \\ \cline{3-4}
\multirow{-2}{*}{\begin{tabular}[c]{@{}l@{}}\textit{Adaptive}\\ \textit{Dataset}\end{tabular}} &
  \multirow{-2}{*}{\begin{tabular}[c]{@{}c@{}}Neural generative\\ model\end{tabular}} &
  Conditional WGAN-GP &
  \begin{tabular}[c]{@{}c@{}}Conditioned adversarial\\ distribution learning\end{tabular} &
  \multirow{-3}{*}{\begin{tabular}[c]{@{}c@{}}Adaptive\\sophisticated attacker\end{tabular}} \\ \hline
\end{tabular}%
}
\end{table*}

%% file: Sections/5-ExperimentalEvaluation.tex
\section{Experimental Evaluation} \label{sec:Evaluation}
This section reports the implementation and measurements obtained from the pipeline in Sec.~\ref{sec:Methodology}, while we provide detailed analysis, interpretation and key takeaways in Sec.~\ref{sec:Discussion}.

\subsection{Experimental Setup}
\textbf{Setup.}
We implement the experiments in Python using \texttt{NumPy}, \texttt{Pandas}, \texttt{scikit-learn}, and \texttt{PyTorch}.
We train classical models (\ac{rf}, \ac{svm}) on CPU,~\footnote{Intel Core i7-12700KF CPU (12 cores, 20 threads, up to 5.0 GHz), 62 GB of RAM.} and neural models (\ac{cnn}, \ac{lstm}, \ac{b-lstm}, \acp{gan}) on one CUDA-enabled GPU.~\footnote{NVIDIA GeForce RTX 3080 Ti GPU (12 GB VRAM, CUDA 12.8).}
Appendix~\ref{sec:ModelsSpec} reports their full specifications.
We split data at the keystroke-session level using an 80\%/20\% train-test ratio.
Each human session and all synthetic counterparts derived from it remain in the same fold, preventing direct leakage of paired sessions between training and testing.

\textbf{Dataset Construction.}
The corpus of \textit{Gonzalez et al.}~\cite{Gonzalez-et-al} contains $18,816$ typing sessions.
Following Sec.~\ref{sec:SyntheticGen}, we preserve each session's \ac{vk} sequence and replace only \ac{ht} and \ac{ft}, aligning human and synthetic sessions while isolating timing from textual structure.
For the \textit{Statistical Dataset}, we use the within-subject synthetic sessions provided by \textit{Gonzalez et al.}~\cite{Gonzalez-et-al}.
The model synthesizing each session is estimated from data belonging to the corresponding human subject.
This grants the attacker subject-specific timing statistics but does not make the detector user-specific.
The detector receives no identity, requires no enrollment, and learns one population-level human-versus-machine boundary.
For the \textit{Adaptive Dataset}, we train both \acp{gan} exclusively on human data and without access to detector parameters.
Each model runs for $20000$ steps with batch size $256$, with checkpoints every $5000$ steps.
We select the checkpoint for dataset generation by assessing Wasserstein-loss stabilization, \ac{ht}/\ac{ft} distributional alignment, variance preservation, and absence of mode collapse.

\subsection{Single-Generator Training Results}
\begin{figure}
    \centering
    \includegraphics[width=\linewidth]{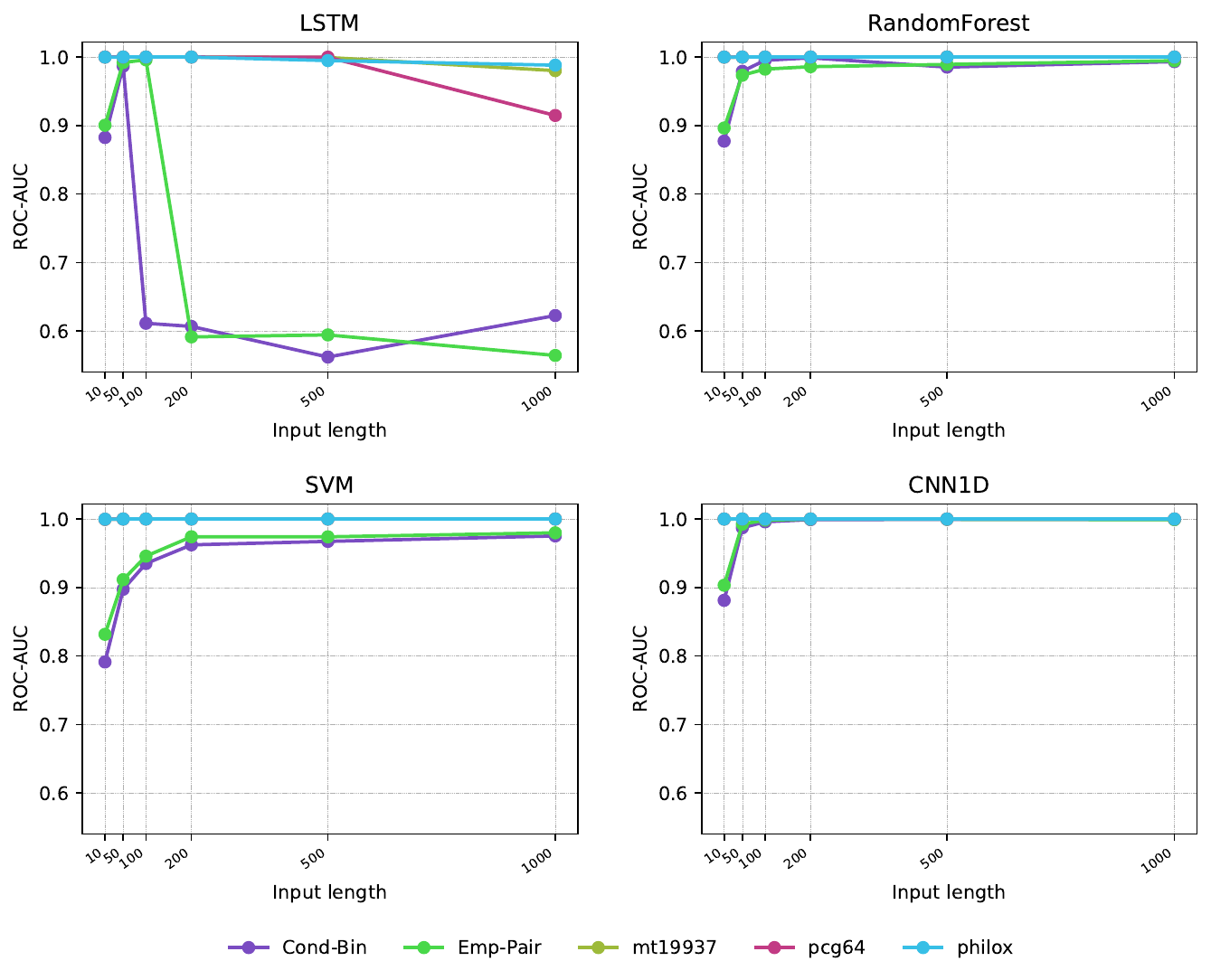}
    \caption{ROC-AUC vs. input length on the Naive Dataset.}
    \label{fig:TrainingNaive}
\end{figure}

\begin{figure}
    \centering
    \includegraphics[width=\linewidth]{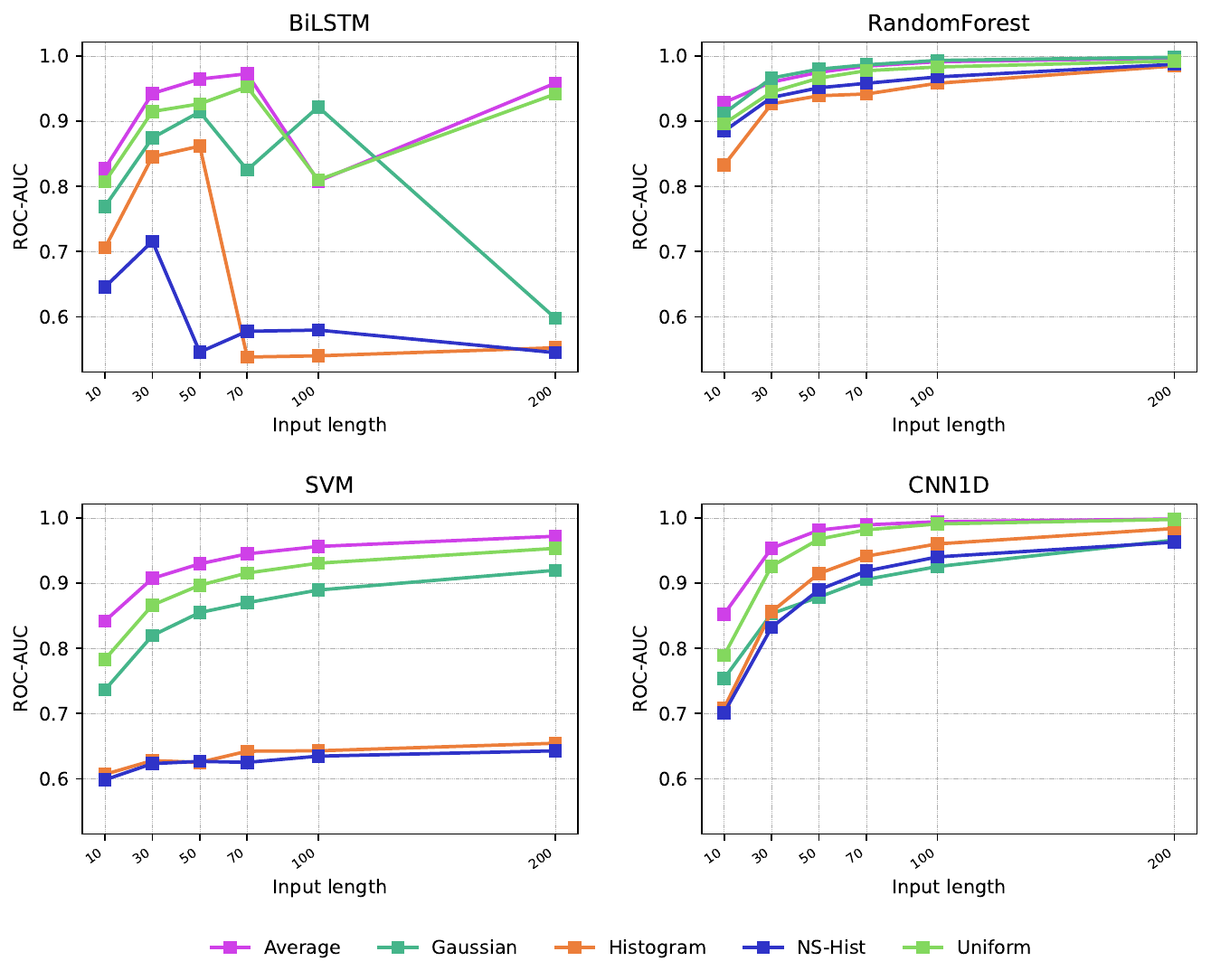}
    \caption{ROC-AUC vs. input length on the Statistical Dataset.}
    \label{fig:TrainingStatistical}
\end{figure}

\begin{figure}
    \centering
    \includegraphics[width=\linewidth]{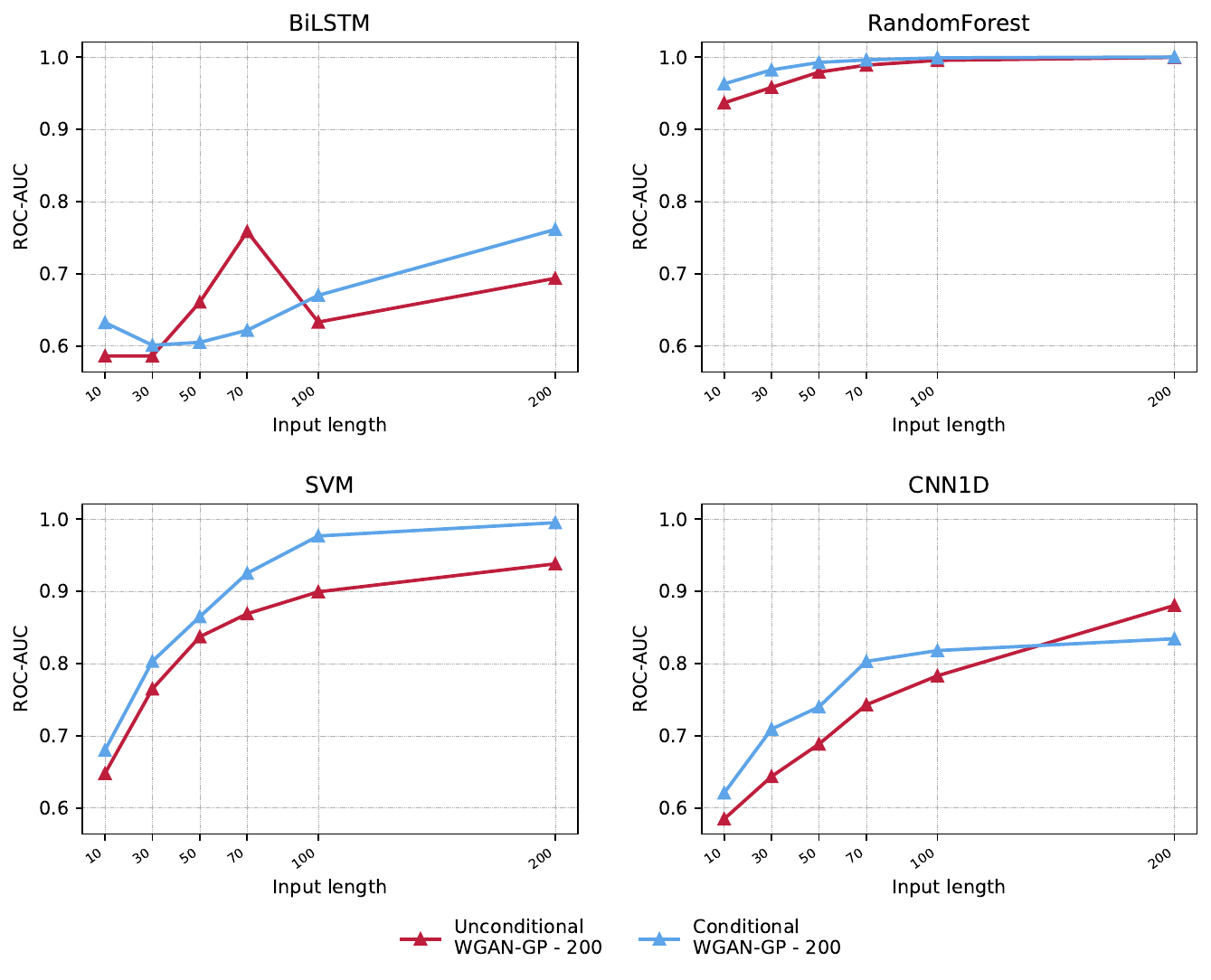}
    \caption{ROC-AUC vs. input length on the Adaptive Dataset.}
    \label{fig:TrainingAdaptive}
\end{figure}

We initially train every detector against individual generators at input lengths of 10, 50, 100, 200, 500, and 1,000 keystrokes.
Fig.~\ref{fig:TrainingNaive} shows that the unidirectional \ac{lstm} degrades as length increases; we therefore use \ac{b-lstm} thereafter.
Performance also saturates beyond moderate lengths, motivating the refined range of 10, 30, 50, 70, 100, and 200 keystrokes used in subsequent experiments.
Figures~\ref{fig:TrainingStatistical} and~\ref{fig:TrainingAdaptive} report the \textit{Statistical} and \textit{Adaptive} results (where 200 denotes the fixed 200-keystroke generation window).
At $70$ keystrokes, \ac{rf} achieves ROC-AUC above $0.9$ for every evaluated generator.
We therefore use $70$ as the representative length for controlled cross-generator analysis, not as an operational threshold or physical payload-interception point.


\begin{figure*}[!h]
    \centering
    \begin{subfigure}[b]{0.45\textwidth}
        \centering
        \includegraphics[width=\linewidth]{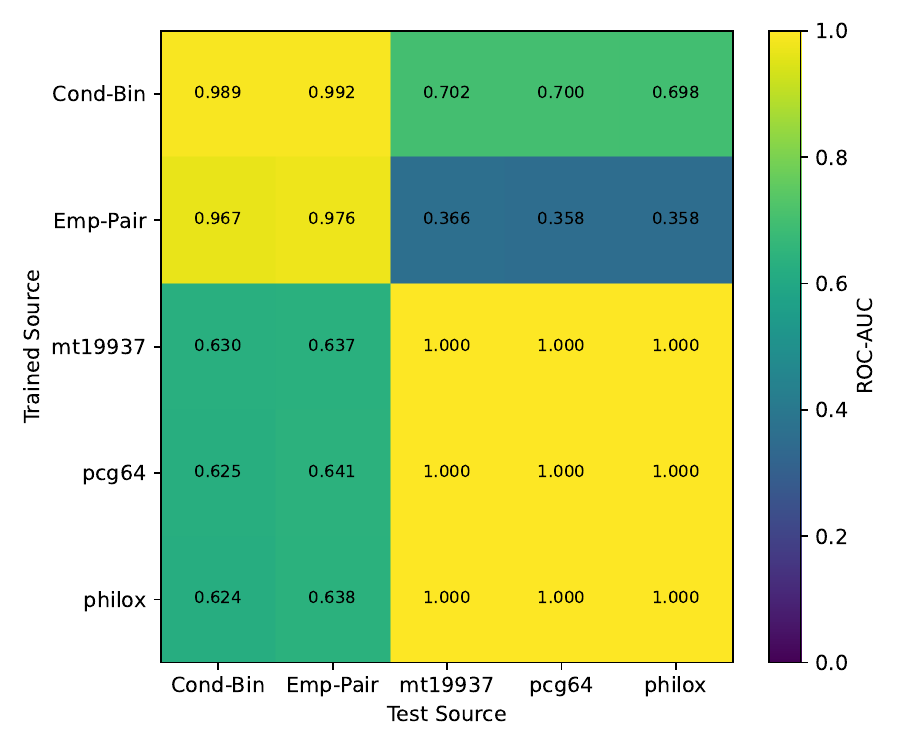}
        \caption{Naive data.}
        \label{fig:SingleGen-Heatmap-RF-70-naive.pdf}
    \end{subfigure}
    \hfill
    \begin{subfigure}[b]{0.45\textwidth}
        \centering
        \includegraphics[width=\linewidth]{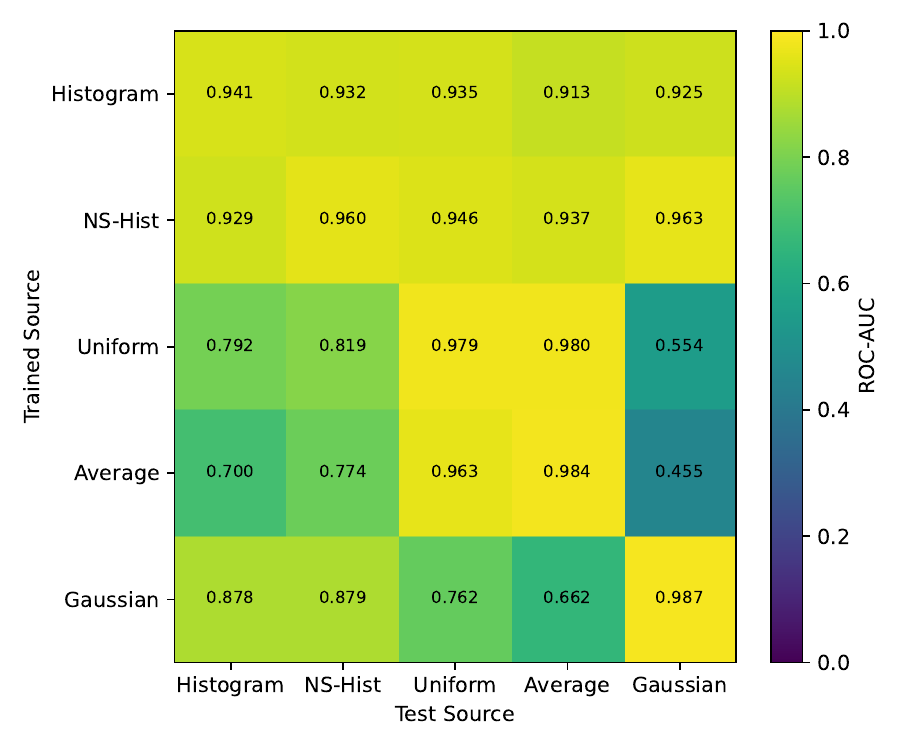}
        \caption{Statistical data.}
        \label{fig:SingleGen-Heatmap-RF-70-statistical.pdf}
    \end{subfigure}
    
    \begin{subfigure}[b]{0.45\textwidth}
        \centering
        \includegraphics[width=\linewidth]{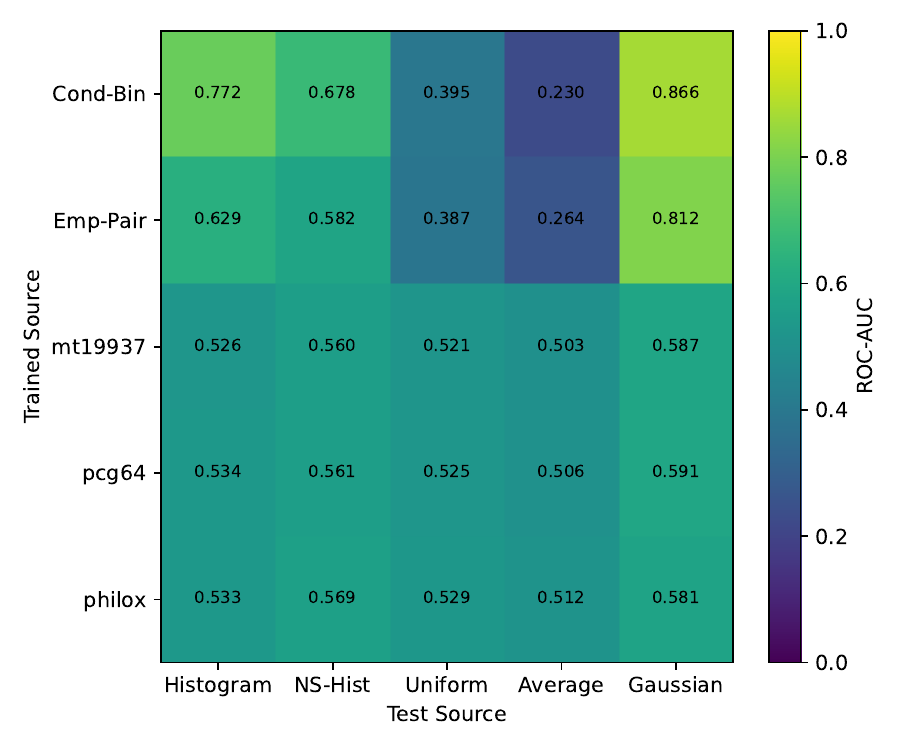}
        \caption{Naive vs Statistical data.}
        \label{fig:SingleGen-Heatmap-RF-70-naive_vs_statistical.pdf}
    \end{subfigure}
    \hfill
    \begin{subfigure}[b]{0.45\textwidth}
        \centering
        \includegraphics[width=\linewidth]{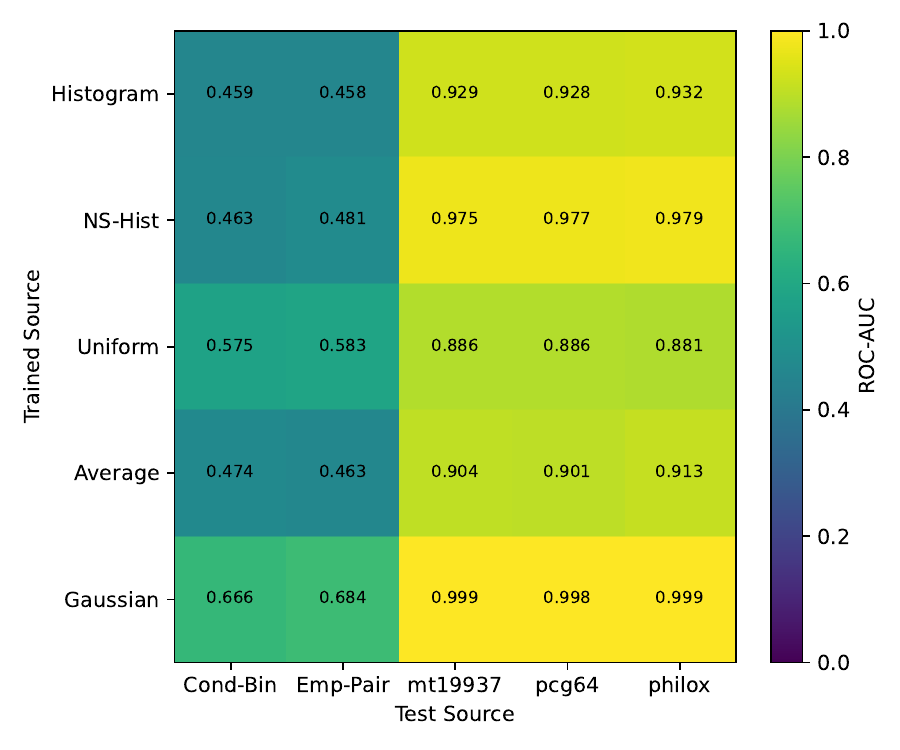}
        \caption{Statistical vs Naive data.}
        \label{fig:SingleGen-Heatmap-RF-70-statistical_vs_naive.pdf}
    \end{subfigure}

    \begin{subfigure}[b]{0.45\textwidth}
        \centering
        \includegraphics[width=\linewidth]{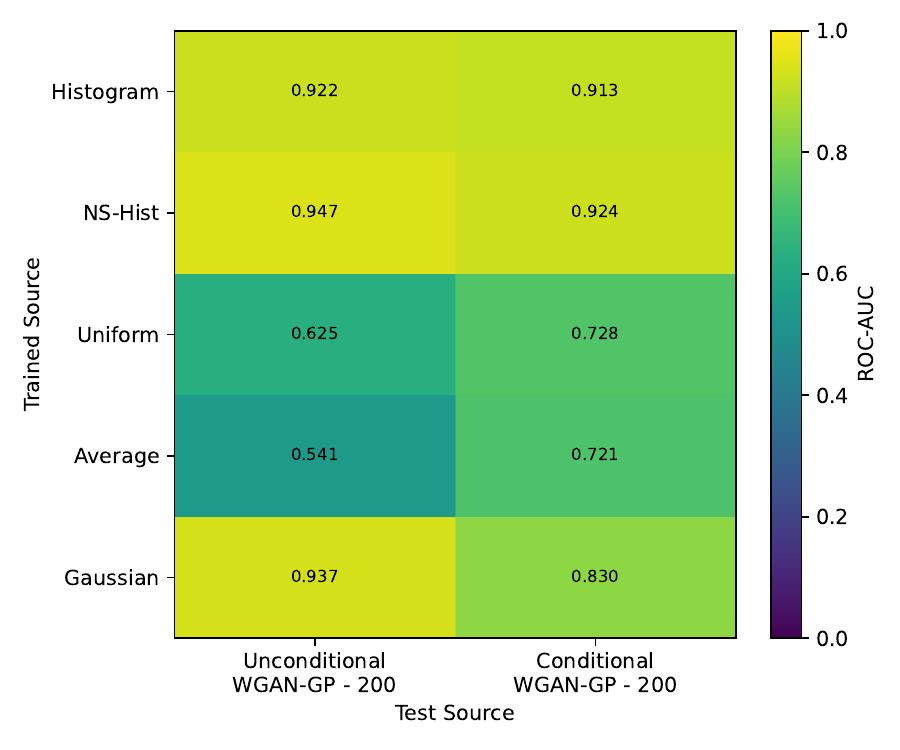}
        \caption{Statistical vs Adaptive data.}
        \label{fig:SingleGen-Heatmap-RF-70-statistical_vs_GANs}
    \end{subfigure}
    \hfill
    \begin{subfigure}[b]{0.45\textwidth}
        \centering
        \includegraphics[width=\linewidth]{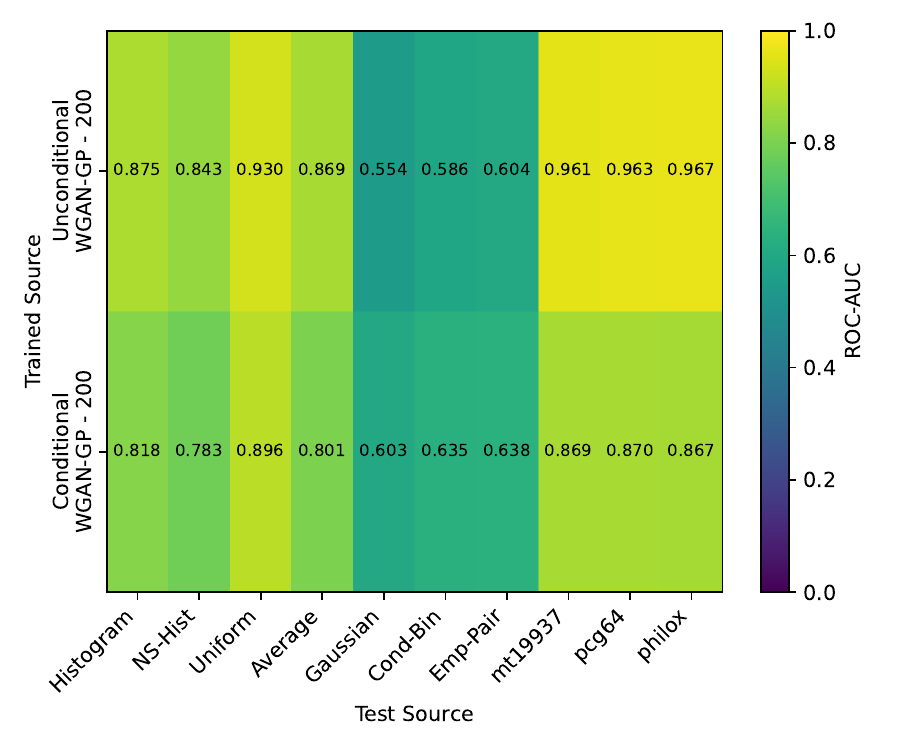}
        \caption{Adaptive vs Naive and Statistical data.}
        \label{fig:SingleGen-Heatmap-RF-70-GANs_vs_All}
    \end{subfigure}

    \caption{Cross-generator ROC-AUC for the \ac{rf} detector trained on single generators at 70 keystrokes.}
    \label{fig:Heatmap-SingleGen}
\end{figure*}

\textbf{Cross-Generator Generalization.}
Fig.~\ref{fig:Heatmap-SingleGen} reports cross-generator results for \ac{rf} at $70$ keystrokes.
Within the \textit{Naive Dataset} (Fig.~\ref{fig:SingleGen-Heatmap-RF-70-naive.pdf}), the \ac{prng} generators transfer strongly among themselves, while \ac{conditional} and \ac{empirical} transfer mutually but poorly across the two clusters.
Within the \textit{Statistical Dataset} (Fig.~\ref{fig:SingleGen-Heatmap-RF-70-statistical.pdf}), \texttt{Histogram} and \ac{non-stationary} transfer consistently, whereas \texttt{Uniform}, \texttt{Average}, and \texttt{Gaussian} show asymmetric or inconsistent transfer.

Cross-family results are also asymmetric.
Statistical-trained detectors transfer effectively to \ac{prng} generators but not to \ac{conditional} and \ac{empirical} (Fig.~\ref{fig:SingleGen-Heatmap-RF-70-statistical_vs_naive.pdf}).
Conversely, Naive-trained detectors generally transfer poorly to Statistical generators (Fig.~\ref{fig:SingleGen-Heatmap-RF-70-naive_vs_statistical.pdf}).
When tested against Adaptive data, \texttt{Histogram}- and \ac{non-stationary}-trained detectors retain strong performance (Fig.~\ref{fig:SingleGen-Heatmap-RF-70-statistical_vs_GANs}).
Conversely, Adaptive-trained detectors do not consistently transfer to simpler Naive and Statistical generators (Fig.~\ref{fig:SingleGen-Heatmap-RF-70-GANs_vs_All}).
These measurements determine the generators considered in mixed training.

\subsection{Mixed-Generator Training Results}
\begin{figure*}[!h]
    \centering
    \begin{subfigure}[b]{0.49\textwidth}
        \centering
        \includegraphics[width=\linewidth]{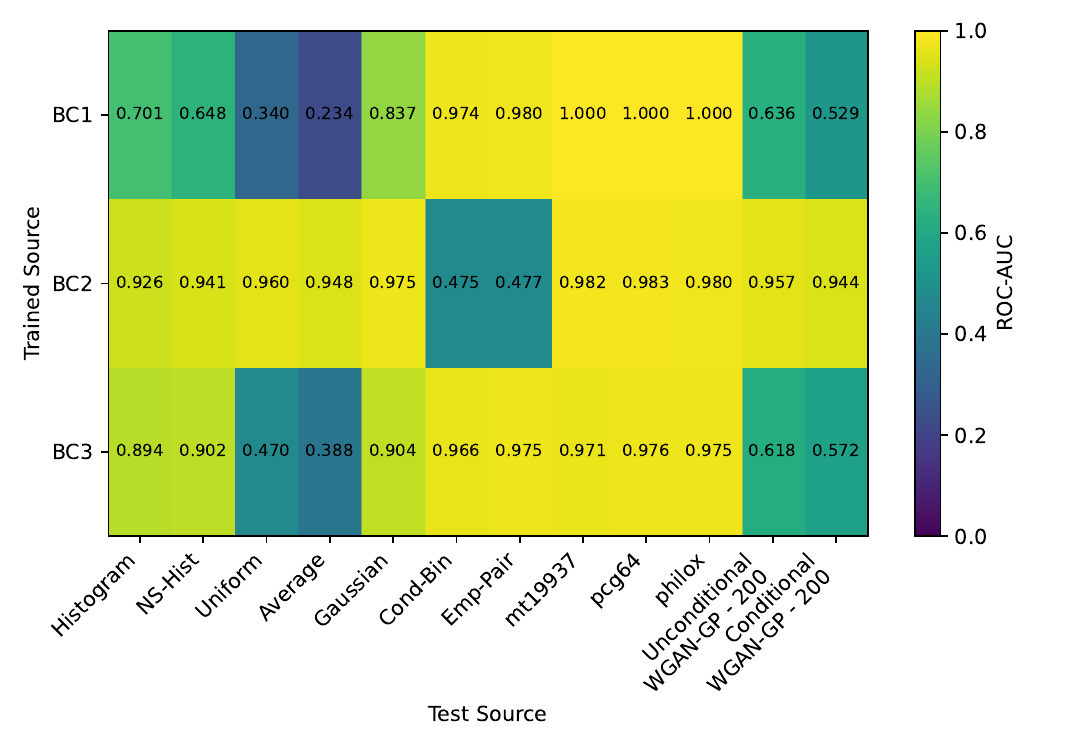}
        \caption{Balanced configurations.}
        \label{fig:MixedGen-Heatmap-RF-70-BC.pdf}
    \end{subfigure}
    \hfill
    \begin{subfigure}[b]{0.49\textwidth}
        \centering
        \includegraphics[width=\linewidth]{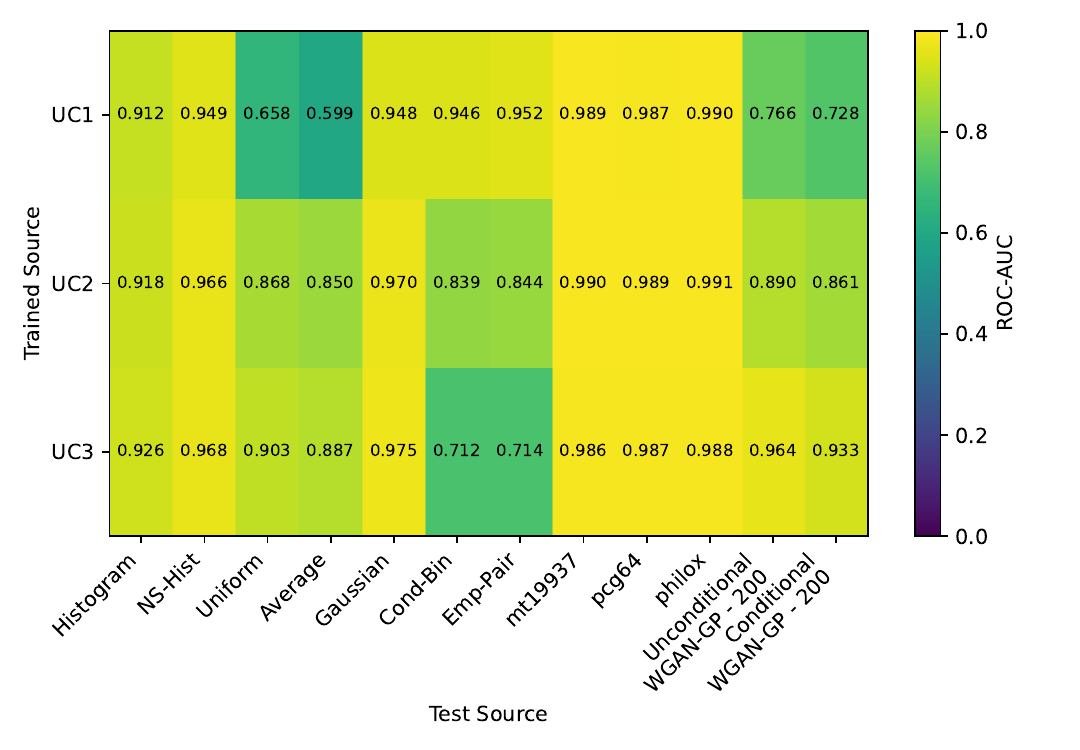}
        \caption{Unbalanced configurations.}
        \label{fig:MixedGen-Heatmap-RF-70-UC.pdf}
    \end{subfigure}
    
    \caption{Mixed-generator training results for the \ac{rf} detector at 70 keystrokes.}
    \label{fig:Mixed-Gen-Results}
\end{figure*}

We combine representative generators selected from the cross-generator measurements.
Table~\ref{tab:TrainingConfigurations} defines the mixtures and their proportions, and Fig.~\ref{fig:Mixed-Gen-Results} reports their performance.

\input{Tables/TrainingConfigurations}

\textbf{Balanced Training.}
BC1 combines complementary Naive generators.
BC2 combines the broadly transferable \ac{non-stationary} generator with \texttt{Uniform} and \texttt{Gaussian}, which perform inconsistently in isolation.
BC3 spans the Naive and Statistical families by combining \ac{conditional}, which transfers to \ac{empirical} but not to \ac{prng} or Statistical generators, with \ac{non-stationary}, which transfers to \ac{prng} and Statistical generators but not to \ac{conditional} or \ac{empirical}.
In Fig.~\ref{fig:MixedGen-Heatmap-RF-70-BC.pdf}, BC1 and BC2 remain stable within their respective families.
BC3 instead degrades on several Statistical generators and \ac{gan}-generated samples relative to BC2.

\textbf{Unbalanced Training.}
The unbalanced configurations emphasize generators with broader transfer (Fig.~\ref{fig:MixedGen-Heatmap-RF-70-UC.pdf}).
UC1 increases \ac{non-stationary} relative to \ac{conditional}, moderately improving Statistical-generator performance while leaving \texttt{Average} and \texttt{Uniform} difficult to separate.
UC2 adds a small \texttt{Average} component, substantially improving both \texttt{Average} and \texttt{Uniform} with limited changes elsewhere.
UC3 adds a small \ac{gan} component, improving Adaptive-generator performance while slightly reducing separability for conditional generators.

\begin{figure}
    \centering
    \includegraphics[width=\linewidth]{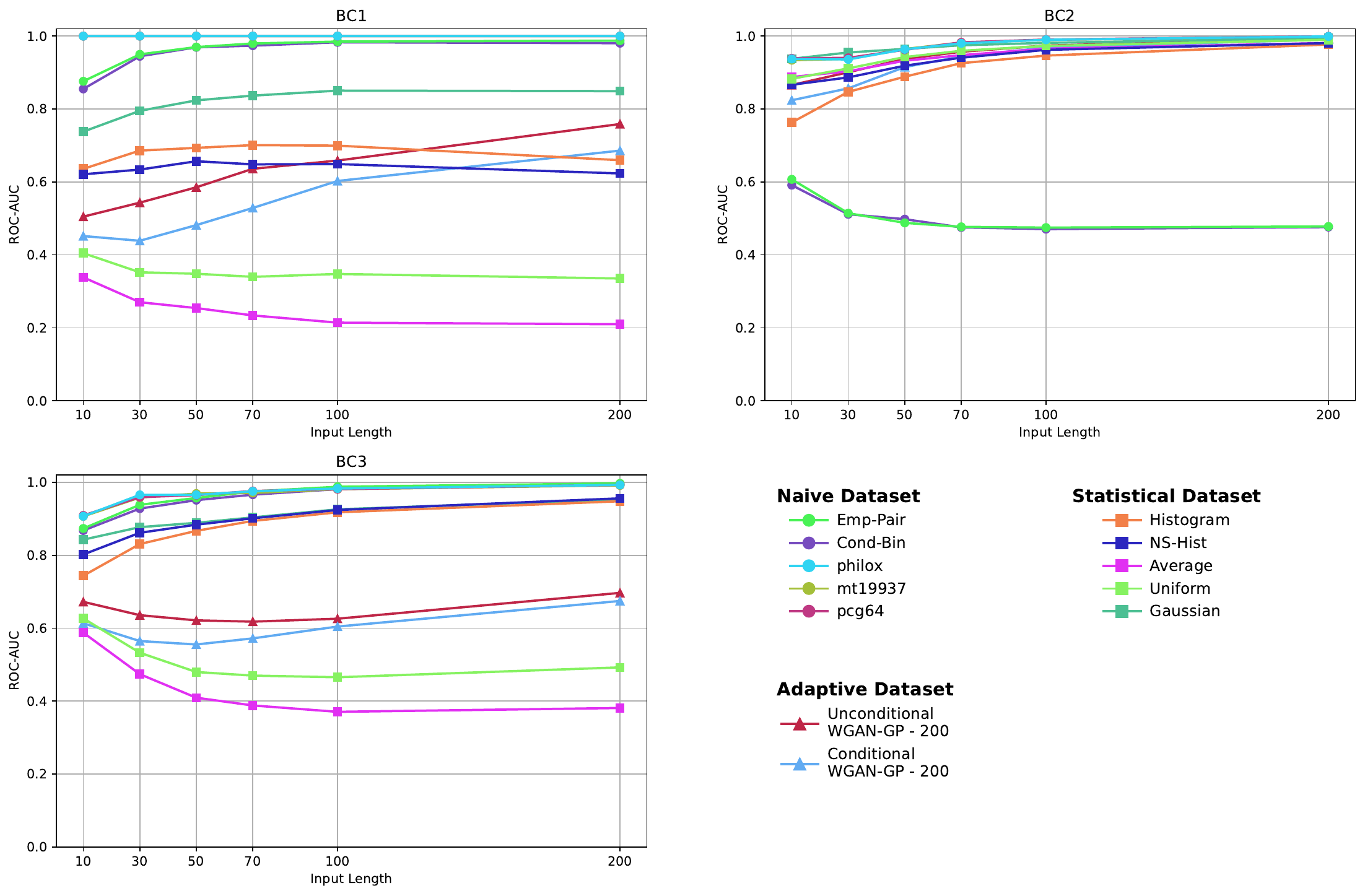}
    \caption{ROC-AUC versus input length for balanced configurations using \ac{rf}.}
    \label{fig:ROC-AUC-input-RF-Balanced}
\end{figure}

\begin{figure}
    \centering
    \includegraphics[width=\linewidth]{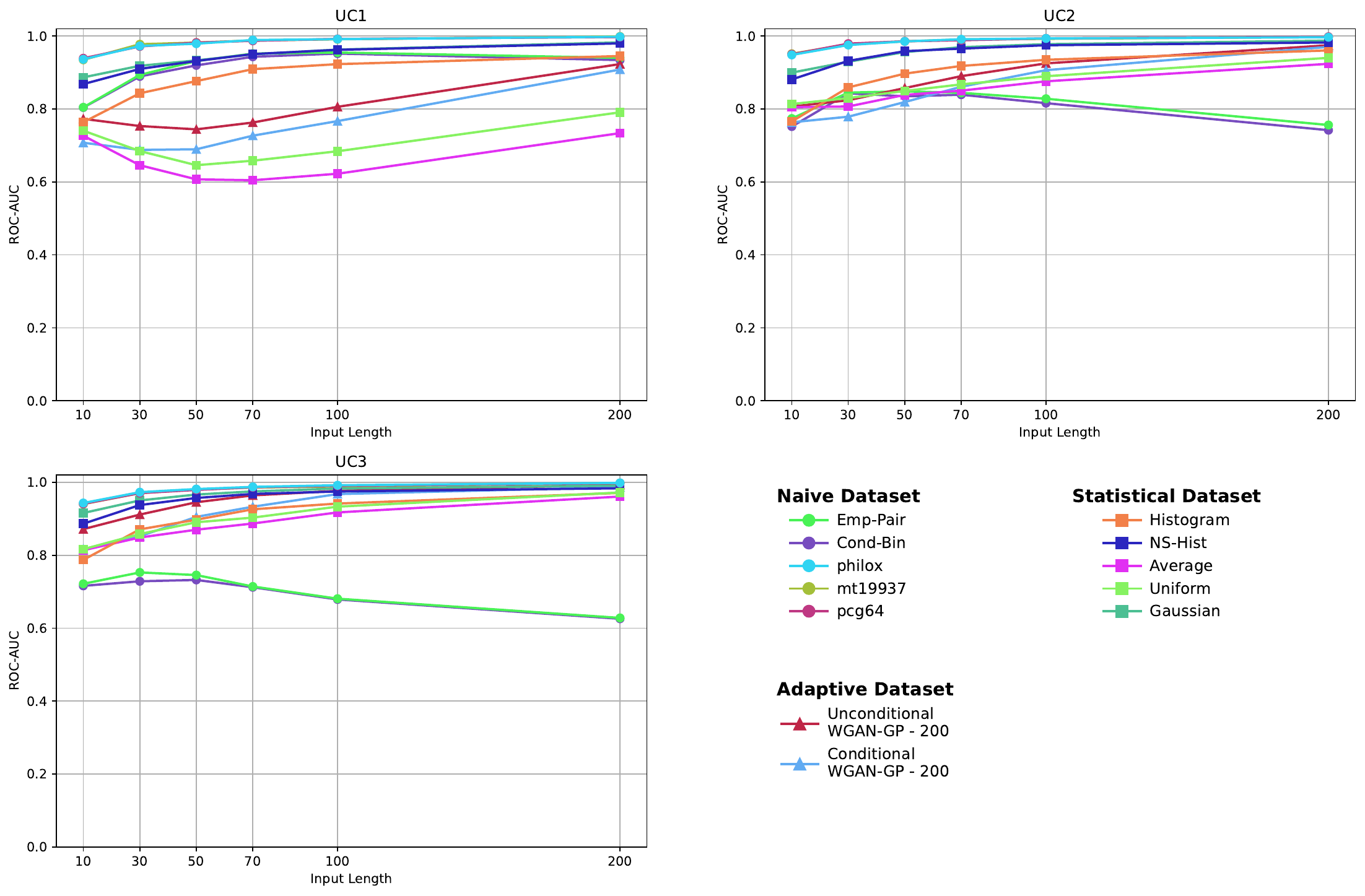}
    \caption{ROC-AUC versus input length for unbalanced configurations using \ac{rf}.}
    \label{fig:ROC-AUC-input-RF-Unbalanced}
\end{figure}

\textbf{Input Size Analysis.}
Figures~\ref{fig:ROC-AUC-input-RF-Balanced} and~\ref{fig:ROC-AUC-input-RF-Unbalanced} report mixed-training performance across input lengths.
Performance generally rises rapidly from $10$ to $70$ keystrokes and stabilizes between $70$ and $100$; longer inputs yield marginal gains concentrated on the most challenging generators.
The pattern holds across balanced and unbalanced configurations and supports using $70$ keystrokes as the representative analysis point.

\subsection{Inference Cost Evaluation}
\begin{figure}[!h]
    \centering
    \begin{subfigure}[b]{0.44\textwidth}
        \centering
        \includegraphics[width=\textwidth]{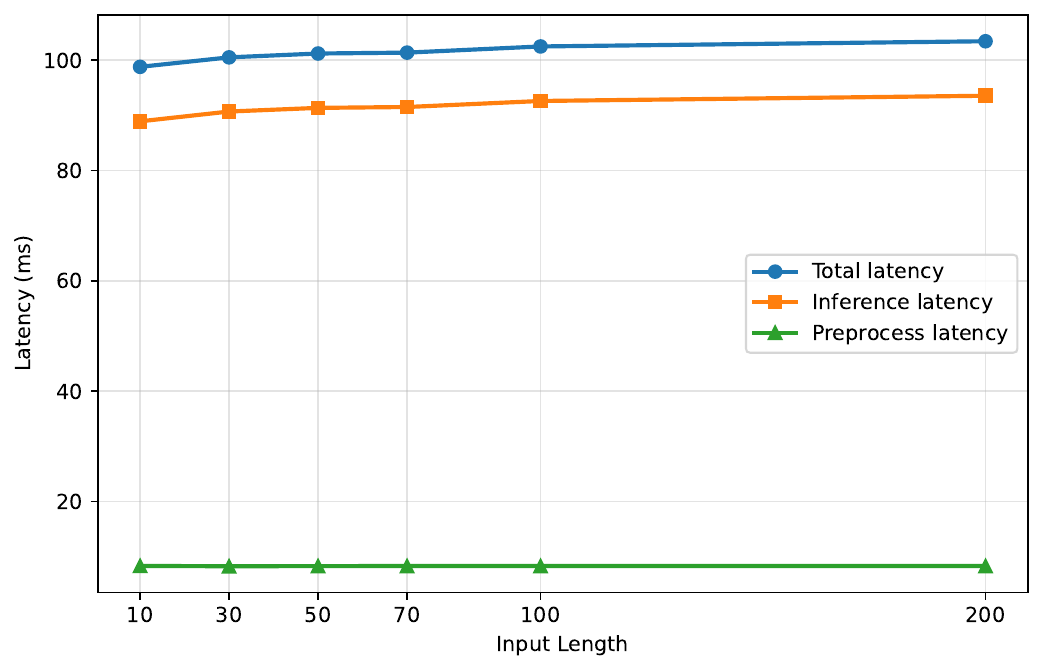}
        \caption{Inference time.}
        \label{fig:InferenceTime}
    \end{subfigure}
    \hfill
    \begin{subfigure}[b]{0.48\textwidth}
        \centering
        \includegraphics[width=\textwidth]{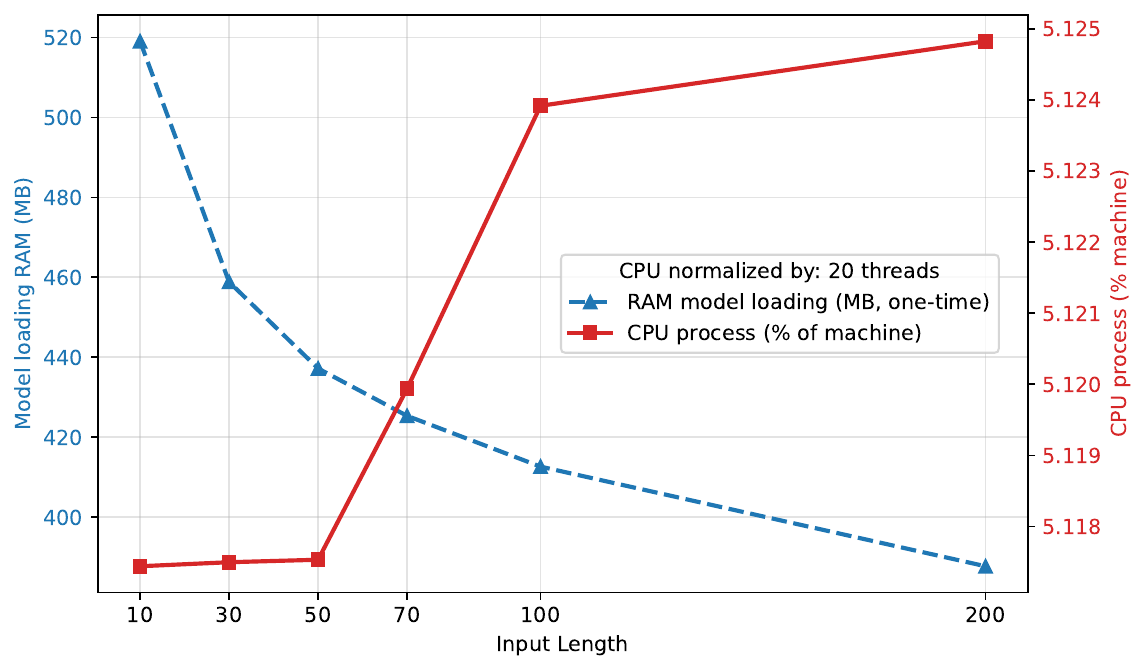}
        \caption{Memory and CPU footprint.}
        \label{fig:MemoryFootprint}
    \end{subfigure}

    \caption{\ac{rf} detector inference-cost analysis.}
    \label{fig:InferenceCost}
\end{figure}

We measure classifier-level single-sample latency, CPU usage, and one-time model-loading memory for the \ac{rf} detector trained with UC3.
For each input length, we record total, preprocessing, and inference latency and capture resource usage through process and device snapshots.
Fig.~\ref{fig:InferenceCost} shows nearly constant preprocessing latency, while inference latency rises only from approximately $89$ to $94$ ms (Fig.~\ref{fig:InferenceTime}); CPU utilization also remains stable (Fig.~\ref{fig:MemoryFootprint}).

The model-loading footprint decreases as input length grows despite increasing input dimensionality.
Because the number of estimators remains fixed at $300$, we hypothesize that longer representations yield trees with fewer effective splits and therefore fewer stored nodes.
Confirming this explanation requires direct comparison of the learned tree topologies.
These platform-specific measurements characterize classifier inference, not the overhead of an end-to-end continuously operating deployment.

%% file: Tables/TrainingConfigurations.tex
\begin{table*}[]
\centering
\renewcommand{\arraystretch}{1.35}
\setlength{\tabcolsep}{5pt}
\caption{Mixed-Generator training configurations.}
\label{tab:TrainingConfigurations}
\resizebox{.7\textwidth}{!}{%
\begin{tabular}{c|c|c}
\hline
\rowcolor[HTML]{C0C0C0} 
    \multicolumn{1}{c|}{\cellcolor[HTML]{C0C0C0}\textbf{\begin{tabular}[c]{@{}c@{}}Training\\ Configuration\end{tabular}}} & \textbf{Included Generators} & \textbf{Representation (\%)} \\ \hline
    
    \textit{BC1} & Cond-Bin, pcg64 & 50 - 50      \\
    \rowcolor[HTML]{EFEFEF} 
    \textit{BC2} & NS-Hist, Uniform, Gaussian           & 33 - 34 - 33 \\
    \textit{BC3} & NS-Hist, Cond-Bin            & 50 - 50      \\
    \rowcolor[HTML]{EFEFEF} 
    \textit{UC1} & NS-Hist, Cond-Bin            & 70 - 30      \\
    \textit{UC2} & NS-Hist, Cond-Bin, Average   & 70 - 25 - 5  \\
    \rowcolor[HTML]{EFEFEF} 
    \textit{UC3} & NS-Hist, Cond-Bin, Unc. \ac{wgan-gp}, Average & 70 - 20 - 7 - 3 \\ \hline
\end{tabular}%
}
\end{table*}

%% file: Sections/6-Discussion.tex
\section{Discussion and Takeaways} \label{sec:Discussion}
In this section, we interpret the evaluation results and provide answers to RQ1 - RQ5.

\textbf{RQ1 -- Lightweight, content-independent keystroke-injection detection.}
All classifiers operate exclusively on \ac{ht} and \ac{ft}, excluding \acp{vk} and reconstructed text.
Thus, classification is content-independent at the detector boundary and requires neither user enrollment nor semantic input features.
Regarding detection feasibility, \ac{rf} detector trained on UC3 achieves ROC-AUC $>0.9$ against every evaluated generator except \texttt{Average} (0.887), \ac{conditional} (0.712) and \ac{empirical} (0.714) (Fig.~\ref{fig:ROC-AUC-input-RF-Unbalanced}).
Fig.~\ref{fig:InferenceCost} also shows a limited and stable classifier-level inference footprint on the evaluated platform.
These measurements establish controlled feasibility of lightweight, timing-only discrimination.
\begin{formal-green}
\textbf{A1.} \textit{Within the evaluated corpus and attacker families, lightweight classifiers can discriminate human from synthetic input using timing alone (i.e., \ac{ht}, \ac{ft}), without per-user enrollment or invasive instrumentation.}
\end{formal-green}

\textbf{RQ2 -- Generalization across attacker models.}
Single-generator training yields near-perfect separability when the generator is known (Figures~\ref{fig:TrainingNaive},~\ref{fig:TrainingStatistical}, and~\ref{fig:TrainingAdaptive}), but cross-generator performance often degrades (Fig.~\ref{fig:Heatmap-SingleGen}).
Closed-world separability is therefore a poor proxy for robustness since many detectors learn generator-specific artifacts rather than invariant differences between human and automated input.
Transfer nevertheless clusters by generative structure.
The \ac{prng}-based generators transfer almost perfectly among themselves; \ac{empirical} and \ac{conditional}, \texttt{Histogram} and \ac{non-stationary}, \texttt{Uniform} and \texttt{Average}, and the two \ac{wgan-gp} variants each exhibit strong mutual transfer.
Hence, the relevant unit of defensive coverage is not necessarily an individual implementation, but the broader mechanism that produces its timing structure.
Within the evaluated families, this supports training on representative generators rather than exhaustively enumerating implementations, shifting the main burden to coverage of structurally distinct classes.
This insight enables a substantial reduction in training requirements while simplifying the overall deployment pipeline.
\begin{formal-green}
\textbf{A2.} \textit{Generalization is driven primarily by exposure to distinct generative structures.
Representative generators can thus provide effective intra-class coverage and reduce training complexity.}
\end{formal-green}

\textbf{RQ3 -- Defender training strategy.}
Mixed-generator training improves overall stability, but its benefits are not monotonic (Fig.~\ref{fig:Mixed-Gen-Results}).
BC3, for example, degrades significantly against \texttt{Average} and \texttt{Uniform} (Fig.~\ref{fig:MixedGen-Heatmap-RF-70-BC.pdf}), although training on \ac{non-stationary} alone nearly perfectly separates both (Fig.~\ref{fig:SingleGen-Heatmap-RF-70-statistical.pdf}).
Under a fixed training budget, distributing samples across heterogeneous generators can dilute the evidence needed to learn a useful boundary for a particular structure.
Conversely, adding underperforming generators in UC1 and UC2 improves robustness to related processes but reduces sensitivity to simpler conditional generators.
More diverse training is therefore not automatically better.
Uniform coverage can exchange generator-specific strength for broader but shallower robustness.
Training composition should instead follow the threat model and the structural coverage identified in RQ2, with augmentation targeted at uncovered or weakly detected classes.
\begin{formal-green}
\textbf{A3.} \textit{Calibrated, threat-model-driven training over representative attacker classes provides the best observed robustness-specificity balance. Thus, augmentation should be selective rather than exhaustive.}
\end{formal-green}

\textbf{RQ4 -- Effect of observation-window length.}
Performance rises rapidly from $10$ to $70$ keystrokes and generally stabilizes between $70$ and $100$. Longer windows provide limited gains for most generators, although some difficult configurations continue to improve.
This is a direct evidence-latency trade-off: short windows permit earlier decisions, whereas long windows provide more temporal evidence.
Because our evaluation is offline and does not replay complete payloads, the keystroke count is only a proxy for decision delay, not elapsed detection time or successful interception.
Accordingly, the observed saturation range is an empirical operating region, not a universal deployment threshold.
\begin{formal-green}
\textbf{A4.} \textit{Larger observation windows do not guarantee commensurate improvements in detection performance, while potentially allowing a greater portion of the malicious payload to be injected before a decision is available.}
\end{formal-green}

\textbf{RQ5 -- Impact of increasing attacker sophistication.}
Greater modeling sophistication does not translate monotonically into stronger evasion or better defensive training data.
Pure \ac{prng} generators are trivially separable, confirming that independent timing samples lack behavioral realism.
The \ac{empirical} and \ac{conditional} generators are harder to detect, especially with short windows, showing that even limited first-order structure can improve evasion.
The trend, however, does not continue with more advanced synthesis.
Detectors trained on the simpler \texttt{Histogram} and \ac{non-stationary} generators transfer to \ac{gan}-generated data with ROC-AUC $>0.9$ (Fig.~\ref{fig:SingleGen-Heatmap-RF-70-statistical_vs_GANs}), whereas detectors trained on \ac{gan} data do not reliably generalize to Naive and Statistical generators (Fig.~\ref{fig:SingleGen-Heatmap-RF-70-GANs_vs_All}).
One plausible interpretation is that the \ac{gan} generators concentrate on a comparatively narrow region of the human-like timing space, while simpler but structurally distinct processes occupy regions not covered by that training distribution.
In conclusion, generator complexity is an unreliable proxy for either evasion strength or defensive coverage, and training against the nominally strongest generator does not reliably generalize to simpler attackers.
\begin{formal-green}
\textbf{A5.} \textit{Evasion depends more on structural diversity than on model complexity; simpler, distinct generators can be as difficult to cover as more sophisticated ones.}
\end{formal-green}

Overall, simple randomization is easy to detect, but additional sophistication yields diminishing returns once the defender covers the relevant attacker classes.
Robustness therefore depends less on detector complexity than on identifying and covering distinct generative structures; conversely, effective evasion requires a structural shift, not merely a more complex synthesizer.

%% file: Sections/6.1-Limitations.tex
\section{Limitations and Future Work} \label{sec:Limitations}
Our conclusions are conditional on the human corpus, attacker generators, and offline evaluation pipeline considered in this study.
Although the human sessions contain measured typing data, the synthetic attacks were not replayed through programmable \ac{hid} hardware and recaptured at the target operating system.
The evaluation therefore omits timing perturbations introduced by device firmware, USB polling, operating-system scheduling, event timestamping, and target load.
These effects may either introduce additional machine-specific signatures or mask artifacts present in the offline generators.
Cross-generator testing reveals some generator-specific overfitting, but cannot guarantee generalization to unmodeled physical devices or attacker implementations.
Accordingly, our results demonstrate controlled feasibility and characterize the factors governing detection, rather than estimate end-to-end deployment performance.

We also do not conduct an online user study.
The public human corpus does not cover every combination of keyboard, operating system, input method, accessibility tool, workload, or transition between human and injected input.
Consequently, our results do not quantify operational false-positive rates or usability impact.
The public corpus and synthetic attacks enable controlled, aligned comparisons across attacker strategies, but cannot replace physical and participant validation.
Future work should replay the evaluated generators through programmable \ac{hid} hardware and recapture their event streams under heterogeneous devices, operating systems, system loads, and timestamping mechanisms.
Participant studies should then assess false positives and usability during realistic interaction, including mixed human-machine input.
Finally, end-to-end experiments should measure elapsed detection latency, the fraction of a payload injected before a decision, and whether detection occurs early enough to support effective intervention.

%% file: Sections/7-Conclusion.tex
\section{Conclusion} \label{sec:Conclusion}
We systematically evaluated timing-only, user-agnostic \ac{hid} injection detection.
In controlled offline experiments, lightweight classifiers using only \ac{ht} and \ac{ft} achieved high discrimination across most evaluated synthetic generators, without per-user enrollment or semantic-text features.
Robustness depended on covering structurally distinct generation strategies: greater synthesis sophistication did not consistently improve evasion, and complex generators did not subsume simpler but structurally different attacks.
Larger observation windows yielded diminishing gains while potentially allowing more of a payload to be injected before detection.
These results establish controlled feasibility and guide future detector design; physical \ac{hid} replay and online user studies remain necessary before deployment-level conclusions.

\par\addvspace{0.5\baselineskip}
\noindent\textbf{Ethical Considerations.}
This study relies on an anonymized dataset of human keystroke dynamics that has been previously collected and released~\cite{Gonzalez-et-al}.
The dataset does not contain Personally Identifiable Information (PII), and no additional data collection involving human participants was conducted.



%% file: Appendix/Specs.tex

\section{Implementation Details}
\label{sec:ModelsSpec}

\textbf{Input representation and sampling.}
The primitive detector observations are \ac{ht} and \ac{ft}, expressed in milliseconds; neither \ac{vk} codes nor reconstructed text are provided to any classifier.
For RF and SVM, the implementation additionally derives
$J_t=|\mathrm{FT}_t-\mathrm{FT}_{t-1}|$, with $J_0=0$, and flattens the resulting $[\mathrm{HT},\mathrm{FT},J]$ window.
$J_t$ is a deterministic transformation of \ac{ft}, rather than an additional measurement or information source, and could therefore be computed by any detector receiving the same \ac{ft} sequence.
The neural models process ordered $[\mathrm{HT},\mathrm{FT}]$ windows directly.
For a session containing at least $L$ events, a contiguous length-$L$ window is sampled uniformly; shorter sessions are padded by repeating their final timing pair.
Neural features are standardized using per-feature statistics computed only from the corresponding detector-training set.

\textbf{Detector models.}
The RF contains $300$ trees, uses all available CPU cores, and otherwise retains the scikit-learn defaults.
The SVM consists of a \texttt{StandardScaler} followed by an RBF-kernel SVC with $C=3$, \texttt{gamma=scale}, and probability calibration enabled.
The LSTM and BiLSTM each contain one recurrent layer with $64$ hidden units; the BiLSTM concatenates the two directions into a $128$-dimensional final-step representation before a scalar linear head.
The CNN contains one same-padded one-dimensional convolution with $64$ output channels and kernel size $5$, followed by ReLU, adaptive max pooling, and a scalar linear head.
All neural detectors are trained for six epochs with batch size $256$, Adam at learning rate $10^{-3}$, and binary cross-entropy with logits; no early-stopping rule is applied.
The split and model seed is $123$.

\textbf{Naive and statistical attacker models.}
The PRNG generators independently draw \ac{ht} and \ac{ft} from uniform intervals bounded by their respective $1$st and $99$th percentiles in a global reservoir of human timing pairs.
The Emp-Pair generator jointly resamples observed $(\mathrm{HT},\mathrm{FT})$ pairs with replacement.
Cond-Bin partitions the preceding \ac{ht} and \ac{ft} into $8\times8$ empirical-quantile bins and samples the next observed pair from the corresponding transition bucket, falling back to the global pair distribution when a bucket is empty.
The Average, Uniform, Gaussian, Histogram, and Non-Stationary Histogram datasets are the \texttt{WithinSubjectAll} outputs released by Gonz\'alez et al.~\cite{Gonzalez-et-al}, rather than local reimplementations of those synthesizers.
During their conversion to the common Parquet representation, negative timings are mapped to zero and values are capped at $1500$~ms.

\textbf{Adaptive attacker models.}
The two adaptive generators are \ac{wgan-gp} models trained only on human windows from the training split.
Before training, positive \ac{ht}/\ac{ft} values are transformed as $\log(\max(x,10^{-3}))$ and standardized per feature using training-human statistics stored in the checkpoint.
The unconditional generator maps a $128$-dimensional Gaussian latent vector to a $128\times L$ representation, applies two residual one-dimensional convolutional blocks, and projects the result to an $L\times2$ timing window.
Each residual block contains two width-$3$ convolutions and Leaky ReLU with slope $0.2$; the final projection is a width-$1$ convolution.
Its critic applies a width-$5$ convolution from two to $128$ channels, Leaky ReLU, two corresponding residual blocks, adaptive average pooling, and a scalar linear head.

The conditional generator uses the same window generator but additionally encodes the preceding $k=10$ timing pairs with two $128$-unit linear layers and concatenates that context representation with the latent vector.
Its critic separately encodes the context and candidate window before concatenating both representations for classification.
Both models are trained for $20{,}000$ generator steps with batch size $256$, five critic updates per generator update, gradient-penalty coefficient $10$, and Adam with $\beta=(0,0.9)$; generator and critic learning rates are $10^{-4}$ and $2\times10^{-4}$, respectively.
Checkpoints are saved every $5{,}000$ steps and selected using the validation criteria described in Sec.~\ref{sec:Evaluation}.
During session synthesis, consecutive generated windows overlap by $25\%$ and are linearly blended before the log transform is inverted.
The conditional stream starts from an all-zero standardized context and subsequently conditions on its most recent generated timings.

\textbf{Splits and evaluation.}
The frozen manifests split the $18{,}816$ aligned session identifiers into $15{,}053$ training and $3{,}763$ test identifiers.
No session identifier occurs in both sets, and each human session and all synthetic counterparts derived from it remain in the same set.
The manifests encode session identifiers rather than user-group assignments; accordingly, they establish session-level and paired-sample separation, but do not by themselves establish user-disjoint folds.
Mixed configurations partition the synthetic identifiers among generators according to the reported BC/UC proportions while retaining balanced human and synthetic classes.
All reported classifiers use synthetic input as the positive class; threshold-dependent metrics use $0.5$, whereas ROC-AUC is computed from continuous scores.

\textbf{Inference-cost instrumentation.}
Figure~\ref{fig:InferenceCost} is based on \texttt{test\_models\_with\_analysis\_cost.py}, applied to the UC3-trained RF detector.
For each input length and test source, the script discards $64$ single-sample warm-up predictions and measures up to $1{,}000$ subsequent predictions.
Preprocessing latency covers session loading and construction of the flattened input; inference latency covers \texttt{predict\_proba}; total latency encloses both stages and the surrounding resource snapshots.
Process CPU use and resident memory are sampled through \texttt{psutil}; model-loading memory is measured in isolated processes over five runs, and the plotting script normalizes process CPU use by the platform's $20$ logical threads before aggregating results by input length.
These measurements characterize classifier-level execution and exclude keystroke capture and continuous endpoint-pipeline overhead.

\textbf{Software environment.}
The released environment pins NumPy~2.4.1, Pandas~3.0.0, scikit-learn~1.8.0, PyArrow~23.0.0, and PyTorch~2.10.0.